\documentclass{article}

\usepackage{blindtext} 
\usepackage[sc]{mathpazo} 
\usepackage[T1]{fontenc} 
\usepackage{microtype} 
\usepackage[english]{babel} 
\usepackage[hmarginratio=1:1,top=32mm,columnsep=20pt]{geometry} 
\usepackage[hang, small,labelfont=bf,up,textfont=it,up]{caption} 
\usepackage{booktabs} 
\usepackage{lettrine} 
\usepackage{enumitem} 
\setlist[itemize]{noitemsep} 
\usepackage{abstract} 
\usepackage{titlesec} 
\renewcommand\thesection{\Roman{section}} 
\renewcommand\thesubsection{\roman{subsection}} 
\titleformat{\section}[block]{\large\scshape\centering}{\thesection.}{1em}{} 
\titleformat{\subsection}[block]{\large}{\thesubsection.}{1em}{} 
\usepackage{titling} 
\usepackage{hyperref} 

\title{Characterization of ISF-VAT performance in weak magnetic nozzle} 
\author{{Satyajit Chowdhury and Igal Kronhaus} \\[1ex] 
\normalsize Aerospace Plasma Laboratory, Faculty of Aerospace Engineering, Technion - Israel Institute of Technology,\\ Haifa, 3200003, Israel; \\ 
\normalsize \href{mailto:satyajit@campus.technion.ac.il}{satyajit@campus.technion.ac.il} 
}

\renewcommand{\maketitlehookd}{%
\begin{abstract}
Vacuum arc thruster performance in a magnetic nozzle configuration is experimentally characterized.  Measurements are performed on a miniature coaxial thruster with an anode inner diameter of $1.8$ mm. The magnetic field B is produced by a single air coil, $18$ mm in diameter. Direct measurement of thrust, mass consumption and arc current are performed. To obtain statistically viable results $\approx 6000$  arc pulses are analyzed at each operational point. Cathode mass erosion is measured using laser profilometry. To sustain thruster operation over several measurement cycles, an active cathode feeding system is used. For $0< B \leq 0.2$ T performance increase over the non-magnetic case is observed with the best thrust to arc power ratio $T/P \approx 9$ $\mu$N/W obtained at $B \approx 0.2$ T. A parametric model is provided that captures the performance enhancement based on beam collimation and acceleration by the magnetic nozzle. For $B > 0.2$ T the arc discharge is shown to be suppressed nullifying any additional gains by the nozzle effect.
\end{abstract}
}
\begin{document}
\maketitle
\section{Introduction}
Vacuum arc thrusters (VAT) are promising propulsion devices for nano-satellites and CubeSats~\cite{levchenko2018,kronhaus2014}, with a few examples experimentally operated in space. VAT are pulsed-dc devices that utilize an arc discharge, across an insulator, between two electrodes to produce thrust. The main advantages of the VAT compared to other electric propulsion devices is its simplicity and scalability to very low power $\sim 1$ W without loss to performance. In a VAT the cathode electrode is consumed as propellant during the discharge. The cathode is eroded in localized regions, where the discharge is attached, known as cathodic spots~\cite{anders2009book}. The metal plasma emitted from these micrometer sized spots is self-consistently accelerated via gas-dynamic expansion~\cite{boxman1996book}. There are two established VAT configuration: coaxial~\cite{zhuang2009} and ring shaped~\cite{kolbeck2019}, termed according to the shape and placement of the cathode with respect to the anode. It is well known that addition of external magnetic field can improve VAT performance~\cite{keidar2005,oleg2017} and the micro-cathode VAT~\cite{keidar2014,zhuang2014}, a ring shaped device, is a well studied case. However, only limited results are documented in the literature for a magnetically enhanced coaxial VATs~\cite{tang2005}. One problem is cathode difficulty in operating such devices over larger number of pulses.

Recently a coaxial VAT with an active feeding mechanism was introduced, known as the inline-screw-feeding VAT (ISF-VAT)~\cite{kronhaus2017,kronhaus2019}. The ISF-VAT enables for the first time a large scale data collection to evaluate the performance of coaxial VATs. Using the ISF-VAT we can introduce the techniques that were previously used in analysis of magnetic nozzle~\cite{gerwin1990} in helicon thrusters~\cite{fruchtman2006,charles2007,takahashi2017,takahashi2019}.
The layout of the paper is as follows: section 1 describes the experimental setup including description of the thruster and thrust measurement; section 2 details the experimental results focusing on thruster performance in varying magnetic induction; section 3 concludes the full experiment. 
 
 \section{Experimental Setup}\label{experimental_setup}

\subsection{Thruster and magnetic setup}
The ISF-VAT is a coaxial vacuum arc thruster with an active feeding mechanism. A central cathode rod is freely disposed within a concentric insulator tube. A second electrode, positioned at the outer edge of the insulator, functions both as the anode of the dc circuit and as the exit plane of the thruster. To keep the VAT geometry constant during long duration operations, the cathode, connected to a metallic headless screw, is advanced at a precise rate inside the insulator in a helical path. The screw provides also an electrical contact with the thruster body that is under negative potential. With the correct selection of the linear advance rate and screw pitch, a balance between cathode erosion and feeding can be achieved. The helical motion both compensates for the radial as well as the azimuthal cathode erosion patterns. This allows for maintaining near constant thruster geometry throughout the operational life of the thruster as well as improved uniformity of the re-coating process, i.e. the process of replenishing the conducting layer on the cathode-insulator-anode interface.

In the present work an ISF-VAT propulsion module (PM) was used. The PM integrates an ISF-VAT with a power processing unit (PPU) and an active feeding system, performing the same set of operations described in Ref.~\cite{kronhaus2017} but in miniaturized form. The PM is being developed at the Aerospace Plasma Laboratory (APL), Technion, and is intended for CubeSat use with a total volume of 2.5 cm $\times$ 9.6 cm $\times$ 9.6 cm and a "wet" mass of $200$ g. A Ti cathode of 0.7 mm in diameter is used in all the tests.

To generate magnetic nozzle, a magnetic system comprising of an air coil wound around the anode was selected. The coil is formed from a 45 turns of 24 AWG copper wire wound around a teflon bobbin,  with an inner diameter of $18$ mm, length of $10$ mm along with a measured inductance $L_{coil}$ $\approx$ 54 $\mu H$ and resistance $R$ $\approx$ 390 m$\Omega$. The magnetic coil is fitted around the ISF-VAT anode and can be relocated along the axis. To achieve maximum magnetic induction, the coil mid length is aligned flush with the cathode surface. The FEMM software~\cite{femm_wiki} was used to calculated the magnetic field topology for a given magnetic coil current $I_{coil}$.

\subsection{Electrical Experimental Setup}

 A 16 V dc laboratory power supply is use to power the PPU dc-to-dc converter that raises the voltage to $\approx40$ V. The PPU is comprised of 3 main components, a capacitor, a discharge coil, and a switch. The PM has a capacitor $C_1$ $\approx$ $0.6$ mF and a ferrite inductor $L_1$ $\approx$ $200$ $\mu H$. An insulated-gate bipolar transistor (IGBT) switch SW-1 is used to charge/discharge the inductor coil. The IGBT can be triggered either by an on board clock or by an external function generator. The triggering signal is a rectangular pulse with amplitude $0 - 10$ V and a frequency $f_{arc}$. The "on" state determines the coil charging duration and is set at 120 $\mu s$. The stored PPU coil energy can be calculated using $\varepsilon_{L_1} = 1/2 L_1 I^2$.

The magnetic coil is powered by a dedicated magnetic circuit. A commercial dc power supply (Lambda, Zup-120) is used as capacitor to discharge through the magnetic coil. It is synchronized with the thruster firing pulses using a separate IGBT switch SW-2 having inverted logic compared to SW-1. The time dependent response of magnetic coil current $I_{coil}$, hence the magnetic induction B, is determined by the power supply voltage $V_B$ and the R-C of the magnetic circuit.

Time dependent measurements of arc current $I_{arc}$ and arc voltage $V_{arc}$ were performed using a Pearson current monitor (Model 150) and a differential high voltage probe. In addition $I_{coil}$ was also measured using a hall current monitor. These signals are simultaneously recorded on a $5$ GS/s oscilloscope.

\subsection{Experimental Setup for Thrust Measurement}

The experiments were performed in APL's vacuum chamber, a cylindrical chamber of $1.2$ m long and $0.6$ m in diameter. Two $700$ l/s turbo-molecular pumps, each backed by $300$ l/m rotary vane pump, are used to maintain a pressure $\sim 10^{-6}$ mbar throughout the test. The chamber pressure is measured by a pfeiffer made compact full Range gauge (model PKR-251). The PM is placed inside the vacuum chamber on a commercial torsion thrust balance~\cite{seifert2013}. The balance was operated in deflection mode with a counter weight on the opposite side of the balance arm. The deflection signal is transmitted optically and recorded on a PC at a sample rate of 2 Hz. Liquid metal contacts are used to decouple the thruster wires from the chamber feedthroughs. the magnetic circuit electronics is placed outside the vacuum chamber. An electrostatic force actuator producing known force was used for thrust calibration. Thrust bias was shown to be negligible at all tested magnetic coil currents. During the experiments video recording of the plume was performed using a 1 MP charged coupled device (CCD) camera set at 60 fps and 1 ms integration time.

\section{Experimental Results}\label{experimental_results}
\subsection{Single Pulse Results}\label{single_pulse_result}
At first the electrical characteristics of single pulse discharge were measured. The $I_{arc}$ and $V_{arc}$ were measured simultaneously with a sampling rate of 2.5 MS/s. The following stages are observed: 1) SW-1 is in "on" state to charge the PPU discharge coil; 2) then SW-1 is set to "off" state, at this moment the rapid change in $dI/dt$ produces a voltage spike reaching $\sim 1$ kV; 3) then an arc is formed $\approx$ 1 $\mu s$ and then starts to decay.

The initial $I_{arc}$ is determined by the coil charging time and is same for every pulse. Therefore, the magnetic energy stored in the coil is also same for each pulse.  However the arc current duration is not constant and is strongly depend on the applied magnetic field. This can be explained by the fact that arc energy is $\varepsilon_{arc}=\int_0^{t_{arc}} I_{arc}V_{arc}dt \sim \varepsilon_{L_1}= const$. Assuming fixed energy, higher arc voltage $V_{arc}$ that is associated with higher magnetic induction limits the arc current duration.

Due to the nature of cathodic arc formation ~\cite{anders2009book}, vacuum arc operation is characterized by statistical fluctuation and therefore fluctuation in arc duration $t_{arc}$ are expected. In order to characterize a mean arc duration in each magnetic case, we have analyzed 1000's of pulses. Using 20 $\mu s$ binning, a Weibull probability distribution function (PDF) was fitted with the data and the mean and error in arc duration are calculated. During arc discharge the $V_{arc}$ is also not constant, notwithstanding a mean value can be calculated.

The time evolution of the $I_{coil}$ during a pulse shows that $I_{coil}$ itself changes during the arc discharge. For comparison, the measured duration of $10\%$ reduction from maximum magnetic induction (from arc ignition). 

The thruster plume luminosity, exhibits significant change in geometry at different magnetic induction values. The plume is hemispherical at $B = 0$ T whereas at $ B \approx 0.12$ T the plume is larger and has a diamond like shape, wider at the base and is pointed in downstream. At the highest magnetic induction tested $B \approx 0.25$ T the plume becomes beam like.

\subsection{Single firing cycle results}

The thruster was operated for a firing cycle duration of $t_{fc} \approx 90$ s at a pulse repetition rate of $f_p = 17$ Hz for a total number of $N_p = f_pt_{fc}\approx 1500$ pulses. Between each firing cycle the cathode was set to advance at an equivalent cathode mass flow rate $\dot{m}_{feed} \approx 2.1$ $\mu g/s$. Both thrust and $I_{arc}$ are measured in each firing cycle.

During these tests the $I_{arc}$ was recorded using an oscilloscope. Due to the limited oscilloscope buffer size, $I_{arc}$ was recorded with a maximum sampling rate of 50 kS/s. This measurement was fully automated using a PC running matlab to acquire $I_{arc}$ data via the oscilloscope and then post process using software edge detect. A valid pulse is considered between currents of 10 - 40 A. This provided us with 10 - 40 samples per pulse depending on arc duration $t_{arc}$. Using trapezoid integration, the software calculates the accumulated charge $Q$ of each valid arc pulse.

Thrust measurements were performed at several magnetic induction values. During the tests, thrust data are continuously recorded at a frequency of 2 Hz. These data are then post processed using matlab. In each firing cycle the thrust data are drift corrected by fitting a linear drift to raw data and then subtracting it. We observe that in all cases thrust is gradually decrease with firing time. This behavior can be explained by gradual erosion of the cathode, since cathode advance is performed only after the firing cycle. It is the reason why cathode feeding is necessary for long duration operation.

Since both the thrust T and $I_{arc}$ are measured independently, it is interesting to correlate them. In order to do so we first calculate $\hat{I}_{arc}$, that is computed by averaging $I_{arc}$ over 0.5 s period, to match the sampling time with thrust. For this specific data set the cross correlation value between T and $\hat{I}_{arc}$ is:

\begin{equation}
\centering
\rho_{T\hat{I}_{arc}} =\frac{\sigma_{T\hat{I}_{arc}}}{\sigma_T\sigma_{\hat{I}_{arc}}}\approx 81\%,
\label{eq:eq_rho_TP}
\end{equation}
where, $\sigma_{T\hat{I}_{arc}}$ is the covariance, $\sigma_T$ and $\sigma_{\hat{I}_{arc}}$ are the thrust and arc current variance.

\subsection{Multiple firing cycle results}

In order to investigate the thruster time-averaged-performance, thrust T, arc power P, and mass consumption were measured in a sequence of at least 4 firing cycles, i.e., for $N_p\geq 6000$. These parameters were then averaged on the entire data set. We observe that the thrust increases significantly for $0 < B < 0.07$ T and reaches a maximum value at $B \approx 0.2$ T, then significantly reduces for $B \approx 0.25$ T. 

The average arc power is calculated according to:
\begin{eqnarray}
\centering
<P(B)> =\frac{\sum\limits_{i=1}^{N_p}Q_i}{t_{fc}}<V_{arc}(B)>,\label{eq:average_arc_power}
\end{eqnarray}
where the mean arc voltage $<V_{arc}>$ is taken. We observe that $<P(B)>$ depends weakly on the magnetic induction. The result is expected as it represents that the PPU output power $\varepsilon_{L_1}f_p=const$ and does not depend on the discharge. The average thrust to power ratio $<T/P>$ can be calculated. At $B = 0$ T, the no coil case is significantly higher in $<T/P>$ with respect to the with-coil case. With application of magnetic induction, $<T/P>$ increases rapidly until doubling the performance at $B = 0.07$ T with respect to no coil case. The maximum is obtained at $B = 0.2$ T. For $B = 0.25$ T the $<T/P>$ is reduced.

The thruster actual mass flow rate $\dot{m}_{meas}$, that is potentially different than the imposed cathode feeding rate $\dot{m}_{feed}$, was also evaluated by measuring the cathode erosion after each test. Precise mass consumption measurements were obtained using laser profilometry of the cathode surface, measuring its shape and depth, before and after test. This technique is described in detail in Ref.~\cite{laterza2019}.

Additional thruster parameters can be calculated, for example the effective specific impulse:
\begin{eqnarray}
\centering
I_{sp} =\frac{<T>}{\dot{m}_{meas}g_0}, \label{eq:specific_impulse}
\end{eqnarray}
where $g_0 = 9.81$ $m/s^2$. We observe that the $I_{sp}$ increases steadily with magnetic induction and that between $0.12 - 0.2$ T the rate of increase is lower. At $0.25$ T there is a significant jump in $I_{sp}$, this is due to lower mass erosion.

The thruster efficiency can be defined as:
\begin{eqnarray}
\centering
\eta_{thruster}=\frac{<T>^2}{2\dot{m}_{meas}<P>}. \label{eq:thruster_efficiency}
\end{eqnarray}
The thruster efficiency is monotonically increasing with magnetic induction, increasing from $\sim 1.1 \%$ at $B = 0$ T to $4.5 \%$ at $B = 0.25$ T.

\section{Conclusions}\label{conclusions}
It was shown that the performance enhancement of a coaxial VAT in magnetic nozzle is substantial up to $8.6$ $\mu N/W$, compared to $4.5$ $\mu N/W$ in the non magnetic case. The improvement in thruster performance between $0 < B < 0.04$ T can be fully attributed to beam collimation. Further improvement, observed at higher magnetic induction values, is attributed to a weak magnetic nozzle effect accelerating the plasma. However, it was also shown that further increase in magnetic induction $B > 0.2$ T causes the $T/P$ to diminish, due to significant decrease in mass consumption at a given arc power.

\section*{Acknowledgments}

We thank Matteo Laterza for his help with the experimental setup. This work was funded by the Israeli Ministry of Science, Technology and Space.



\end{document}